\documentclass[runningheads]{llncs}
\usepackage{multirow} 
\usepackage{graphicx}
\usepackage{subcaption}
\usepackage{algorithm}
\usepackage{algpseudocode}
\usepackage{amsmath}
\usepackage{comment}

\usepackage[T1]{fontenc}
\usepackage{graphicx,verbatim}
\usepackage{color}
\begin{document}
\definecolor{addColor}{rgb}{0.02,0.55,0.0}
\newcommand{\add}[1]{\color{black}#1}
\title{Uncertainty-driven training for three-dimensional calibrated lung nodule classification}
\titlerunning{Uncertainty-driven training for lung nodule classification}
%
\author{Giuseppe Tripodi \and
Alessandro De Rosis \and
Saleh Rezaeiravesh}
\authorrunning{G. Tripodi et al.}
%
\institute{School of Engineering, The University of Manchester, Manchester (UK)
\email{giuseppe.tripodi@postgrad.manchester.ac.uk}\\}

\maketitle              
\begin{abstract}
In this work, we present an uncertainty‑driven training framework for three‑dimensional computed tomography \add{(CT)} lung nodule classification, where validation‑based uncertainty \add{estimates guide} loss reweighting to \add{enhance predictive performance} and probability calibration.
Two Uncertainty Quantification (UQ) methods are considered: Monte Carlo Dropout (MCD) and Evidential Deep Learning (EDL). Both provide per‑class uncertainty estimates that modulate the loss and encourage focus on hard or unreliable classes. 
\add{The framework is evaluated with ResNet, DenseNet, EfficientNet, Vision Transformer (ViT), and Swin Transformer backbones on two datasets: the clinical LIDC‑IDRI cohort and the NoduleMNIST3D benchmark. }
Uncertainty‑driven training achieves classification performance similar to conventional training while substantially improving calibration, with an expected calibration error (ECE) reduced by up to $65\%$ on LIDC‑IDRI. EDL attains competitive performance on shallower architectures with single‑pass inference, whereas MCD is more robust on deeper networks.
\add{Analysis across architectural families reveals that uncertainty-driven training benefits convolutional backbones more consistently than transformer-based architectures: EDL in particular degrades on ViT, suggesting that the Dirichlet evidence parameterisation may interact unfavourably with attention-based architectures at lower input resolutions. 
A posteriori temperature scaling proves highly effective across all configurations, indicating  that a simple scalar calibration can be competitive even without explicit uncertainty-aware training.}
Our results indicate that integrating UQ into the training loop can significantly improve probabilistic calibration and support more trustworthy deployment of three-dimensional medical imaging models.


\end{abstract}

\section{Introduction}

\begin{figure}[t]
    \centering
    \includegraphics[width=1\linewidth]{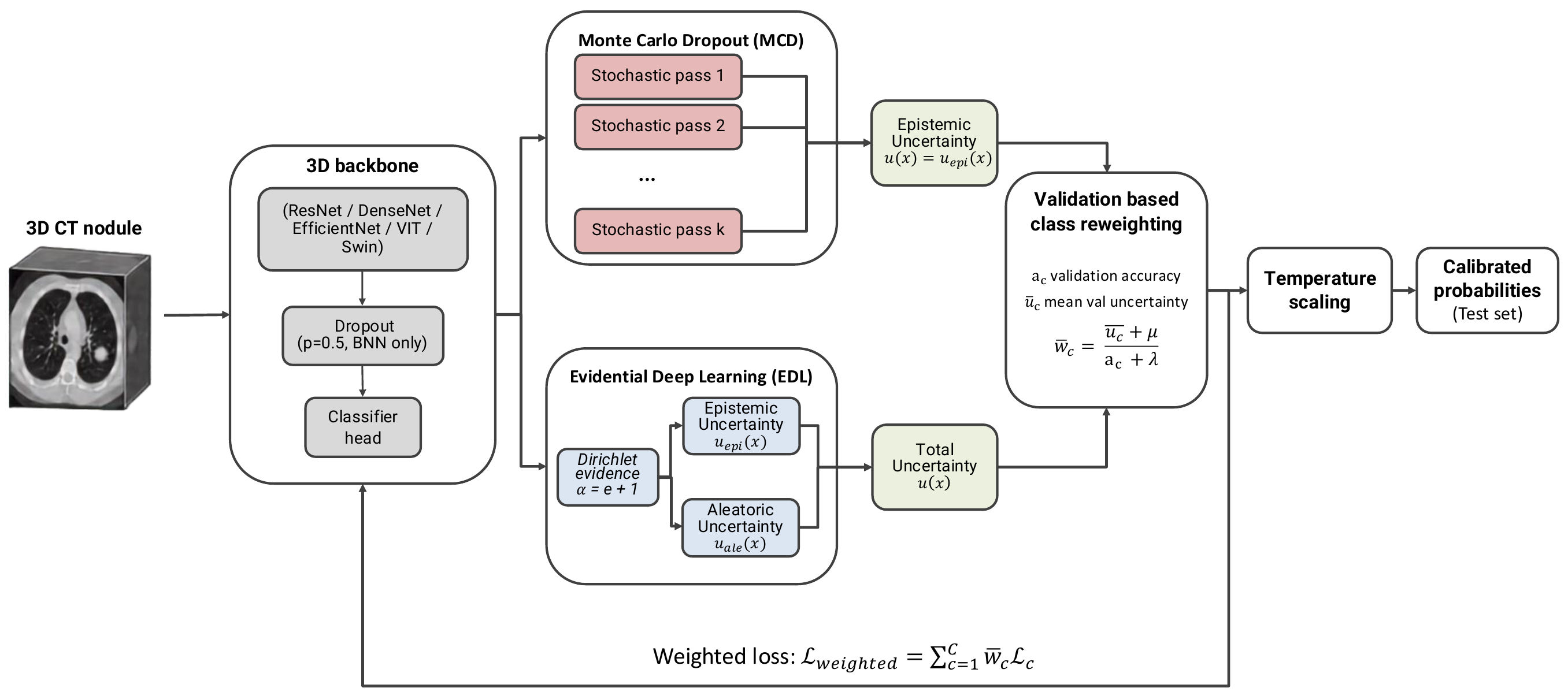}
    \caption{Overview of the proposed uncertainty-driven training framework for lung nodule classification. Starting from preprocessed 3D CT nodule patches, multiple 3D backbones are trained using the MCD and EDL methods. Epistemic and aleatoric uncertainties estimated on the validation set drive class reweighting and a weighted loss, while \textit{a posteriori} temperature scaling applied to validation logits yields calibrated test-time probabilities for malignancy prediction.}
    \label{fig:workflow}
\end{figure}

Lung cancer ranks among the leading causes of oncological mortality worldwide, with thousands of deaths reported annually in the United Kingdom~\cite{Bray2024GLOBOCAN}.
A timely prediction of pulmonary nodules plays a significant role in the early detection of lung cancer. However, a reliable distinction between benign and malignant nodules represents a relevant challenge for contemporary computed tomography (CT)~\cite{GU2021101886}.
Deep learning methods have achieved strong performance for nodule detection and classification, yet they often produce overconfident predicted probabilities~\cite{guo2017calibrationmodernneuralnetworks,Zahari2023,GU2021104806}.
In safety-critical decision-support settings, reliable uncertainty estimates are essential. Furthermore, well-calibrated models enable clinicians to identify cases \add{with high predictive uncertainty}, prioritise expert review, and communicate risk more transparently to patients~\cite{sensoy2018,gawlikowski2022surveyuncertaintydeepneural}.
\\\indent 
Many efforts on \add{Uncertainty Quantification} (UQ) in medical imaging have \add{been} largely focused on \textit{a posteriori} analyses and pixel-wise uncertainty maps. UQ estimators as Monte Carlo~\add{(MC) methods} and related Bayesian approximations, are commonly adopted to assess prediction reliability, filter low-confidence cases, \add{and} separate certain from uncertain predictions in lung cancer and skin lesion analysis~\cite{Zahari2023,zahari2024,enhancing_nodule_risk_uncertainty,ABDAR2021243,ABDAR2021104418}.
Recently, several authors have interpreted uncertainty driven training as a form of cost‑sensitive learning~\cite{Elkan2001}, where unequal misclassification costs are encoded as class‑dependent weights so that errors on clinically high‑risk or under‑represented classes are penalised more heavily~\cite{KRAWCZYK2014554}. 
Dawood~\emph{et al.}~\cite{Dawood_2023} used pixel-wise uncertainty to weight the loss in magnetic resonance (MR) segmentation. Zheng~\emph{et al.}~\cite{zheng2022uncertainty} incorporated uncertainty maps into the loss to highlight informative regions, and Yu~\emph{et al.}~\cite{https://doi.org/10.48550/arxiv.1907.07034} used uncertainty to weight the consistency loss of a teacher-student model to do semi‑supervised \add{three-dimensional} (3D) left atrium segmentation.
\newline \indent 
In the present study, we introduce an \emph{uncertainty-driven training framework} for 3D lung nodule classification that integrates \add{UQ} directly into the optimisation process.
The core idea is to use~\emph{per-class, validation-based} uncertainty as a cost signal to reweight classes during training, explicitly targeting both class imbalance and model calibration.
To avoid circular dependencies, uncertainty is computed only on a held-out validation set that is never used for gradient updates. After an initial warm-up phase, class weights are updated at fixed epoch intervals so that classes that are both underperforming and highly uncertain are upweighted in the loss.
Our framework is instantiated with two complementary UQ estimators: Monte Carlo Dropout (MCD), which estimates epistemic (model-related) uncertainty via multiple stochastic forward passes~\cite{gal2016}, and Evidential Deep Learning (EDL), which computes epistemic and aleatoric (data-related) uncertainties via a single pass using a Dirichlet distribution as class probabilities~\cite{sensoy2018}. 
After training, we further apply an \textit{a posteriori} temperature scaling to improve calibration of the final predictions.
\add{Figure~\ref{fig:workflow} illustrates the overall framework pipeline, while the underlying methodologies are detailed in Section~\ref{sec:meth}.}
%
Our approach is evaluated against two  3D CT datasets, LIDC-IDRI~\cite{armato2015lidc} and NoduleMNIST3D~\cite{Yang_2023}, using 
\add{multiple convolutional and transformer-based backbones.}
\add{Our numerical experiments show that incorporating UQ into the training loop improves model calibration and returns uncertainty estimates that are positively correlated with prediction errors. On average, misclassified samples tend to receive higher uncertainty scores than correctly classified ones, although the effect varies across architectures. 
At the same time, classification performance remains comparable to conventional training, indicating that incorporating UQ into the training process enhances the reliability of the 3D lung nodule classifier without sacrificing accuracy. 
}

\section{Methodology}\label{sec:meth}

\subsection{Data description and preprocessing}
We evaluate our \add{framework} on two 3D lung nodule datasets: LIDC-IDRI~\cite{armato2015lidc} and NoduleMNIST3D from MedMNISTv2~\cite{Yang_2023}.
For LIDC-IDRI, we follow the protocol by Zhao~\emph{et al.}~\cite{Zhao2020}, where CT scans with a slice thickness greater than~$3\,\mathrm{mm}$ are excluded and only nodules annotated by all four radiologists and with a diameter greater than or equal to $3\,\mathrm{mm}$ are considered.  
Ground-truth malignancy labels are derived by averaging the radiologists' malignancy scores; nodules with an average score below~$2.5$ are labelled as benign, whereas those with an average score above~ $3.5$ are labelled as malignant. 
\add{Nodules in the ambiguous range~$(2.5, 3.5)$ are excluded to reduce label noise.}
After this filtering,~$809$ nodules remain, comprising~$428$ benign~($52\%$) and~$381$ malignant~($48\%$).
For each nodule, we extract a $48\times48\times48$ voxel patch centred on the annotated coordinates and then apply a min–max normalisation. 
To improve generalisation, we use 3D data augmentation during training, including random cropping, rotation, flipping, and intensity adjustments (gamma, brightness, contrast).
NoduleMNIST3D contains 1633 3D CT nodules (1232 benign, 401 malignant) with an original resolution of $28\times28\times28$ voxels~\cite{Yang_2023}. These volumes are resized to $64\times64\times64$, independently min–max normalised, and augmented with the same 3D transformations as the LIDC-IDRI.
\subsection{Model architectures and training strategies}

\add{Five families of 3D backbones are evaluated: ResNet (ResNet-18, ResNet-50)~\cite{ResNet}, DenseNet (DenseNet-121, DenseNet-169)~\cite{DenseNet}, EfficientNet~\cite{EfficientNet}, Vision Transformer (ViT3D)~\cite{ViT}, and Swin Transformer (SwinViT3D)~\cite{SwinTranformer}. 
Each backbone is instantiated in two configurations. 
The \emph{standard} variant disables dropout at inference and produces a single deterministic softmax output. 
The Bayesian Neural Network (BNN) variant introduces a single dropout layer (with probability $p=0.5$) directly before the classification head. During the inference, the network is kept in training mode so that dropout remains active, while all batch-normalisation layers are switched to evaluation mode to preserve the running statistics learned during training. 
Following Gal and Ghahramani~\cite{gal2016}, repeated stochastic forward passes yield a distribution over class probabilities from which epistemic uncertainty can be estimated.
When the evidential objective is selected, the BNN backbone requires no further structural changes: the output of the final linear layer is passed through a ReLU activation to yield non-negative evidence, without applying the softmax. 
This shared design means that each backbone can be trained using either MCD or EDL, with the training objective as the only difference.}
\add{The following four training strategies are compared.}
\add{\paragraph{Focal loss baseline (FL).} Models are trained with a focal loss~\cite{lin2018focallossdenseobject} (class-balancing parameter $\alpha = 0.25$ and focusing parameter $\gamma = 2$) and uniform class weights:
\begin{equation}
    \mathcal{L}_{\text{FL}} = -\alpha \sum_{c=1}^{C} (1 - p_c)^\gamma \log p_c.
    \label{eq:focal}
\end{equation}
Here, no uncertainty information is used during training.} 
%
\add{\paragraph{MC Dropout-guided reweighting (MCD).} At the end of each update cycle, $K = 20$ stochastic forward passes are run on the validation set to estimate variance~$\operatorname{Var}_t[\cdot]$. Then, per-sample epistemic uncertainty is defined as, 
\begin{equation}
    u(x) = \sqrt{\sum_{c=1}^{C} \operatorname{Var}_t \!\left[ p^{(t)}(c \mid x) \right]},
    \label{eq:mcd_unc}
\end{equation}
and averaged \add{over the validation set} within each class to obtain $\bar{u}_c$. Class weights are then set proportionally to both the inverse per-class validation accuracy $a_c$ and the mean uncertainty $\bar{u}_c$:
\begin{equation}
    \bar{w}_c = \frac{\bar{u}_c + \mu}{a_c + \lambda}\,,
    \qquad
    w_c = \operatorname{clip}\!\left( \frac{\bar{w}_c}{\max_k \bar{w}_k} \cdot \gamma,\; w_{\min}, w_{\max} \right)\,,
    \label{eq:weights}
\end{equation}
with $\gamma = 2$, $w_{\min} = 0.5$, $w_{\max} = 2$, and smoothing constants $\mu$ and $\lambda$. Weights are activated after a warm-up period and updated periodically. Deriving them from the validation set prevents circular dependency with the training objective.
}
\add{\paragraph{Evidential Deep Learning (EDL).} The classification head is replaced by a ReLU-activated evidence head, yielding Dirichlet parameters $\alpha_c = e_c + 1$~\cite{sensoy2018}, where~$e_c$ is the evidence assigned to class~$c$. 
The total evidence $S=\sum_{c=1}^C \alpha_c$ defines the evidential uncertainty $u(x)=C/S$, where low evidence corresponds to high predictive uncertainty. 
Training minimises a loss with two parts: a Bayesian risk term that penalises misclassification under the Dirichlet posterior, and a Kullback-Leibler (KL) divergence term $ D_{\text{KL}}$ that regularises the evidence towards a uniform Dirichlet prior, $ \operatorname{Dir}(\mathbf{1})$~\cite{sensoy2018}:
\begin{equation}
    \mathcal{L}_{\text{EDL}} = 
    \sum_{c=1}^{C} \left[\left(y_c - \mu_c\right)^2 + \frac{\mu_c(1-\mu_c)}{S+1}\right]
    + 
    \lambda_{reg}\, D_{\text{KL}}\!\left[ \operatorname{Dir}(\tilde{\boldsymbol{\alpha}}) \,\|\, \operatorname{Dir}(\mathbf{{\boldsymbol{1}}}) \right]\,,
    \label{eq:edl}
\end{equation}
where $\mu_c = \alpha_c/S$, $y_c$ is the one-hot label value, and $\lambda_{reg}$ is an annealing schedule. Class reweighting follows the same scheme as MCD (Eq.~\ref{eq:weights}), but the total per-class uncertainty $\bar{u}_c $ is computed from a single validation pass.
}
\add{\paragraph{Calibration-aware training (CA).}
Focal loss is augmented with a differentiable \add{surrogate for Expected Calibration Error} (ECE) that replaces hard bin assignments with Gaussian soft weights~\cite{karandikar2021soft}:
\begin{equation}
    \mathcal{L}_{\text{CA}} = \mathcal{L}_{\text{FL}} + \mu_{CA}\, \widehat{\operatorname{ECE}}_{\text{soft}},
    \label{eq:ca}
\end{equation}
where $\mu_{CA}$ balances the accuracy–calibration trade-off. This directly penalises miscalibration during training without any explicit uncertainty estimation.}
\subsection{Experimental setup and evaluation}
\add{All models are implemented in PyTorch~\cite{https://doi.org/10.48550/arxiv.1912.01703} and trained on an NVIDIA A100 GPU using a fixed stratified 70:15:15 training, test and validation split performed at the patient level to prevent data leakage. 
The uncertainty-based class weights are derived exclusively from the validation set and are not updated using training samples, avoiding circular dependencies with the optimisation objective.
All models are optimised with AdamW~\cite{kingma2017adammethodstochasticoptimization} for up to 300 epochs with a cosine-annealing learning-rate schedule, a batch size of 16, and early stopping. 
%
For focal loss we set $\gamma = 2$ and $\alpha = 0.25$ following Zahari~\emph{et al.}~\cite{Zahari2023}. The EDL regularisation coefficient $\lambda_{\text{reg}}$ is tuned per architecture and ranges from~$0.03$ to~$0.3$, with shallower architectures generally requiring larger values to prevent evidence collapse. For CA, the ECE penalty weight is set to $\mu_{CA} = 0.03$.
}
\add{The hyperparameters in the uncertainty-driven weighting scheme and the calibration-aware loss were selected through systematic grid searches.}
Accuracy (ACC), precision (PRE), recall (REC), F1~score, and area under the ROC curve (AUC) are used as the classification performance metrics~\cite{Hicks2022}.
Calibration is assessed using ECE, Maximum Calibration Error (MCE)~\cite{guo2017calibrationmodernneuralnetworks}, and Brier Score~(BS)~\cite{Dawood_2023}.
\add{ The ability of the model to identify its own mistakes is evaluated via the Area Under the ROC Curve for Error Detection (AUROC-ED)~\cite{https://doi.org/10.48550/arxiv.1610.02136}, which treats misclassified samples as the positive class and ranks them by epistemic uncertainty.
Higher value indicates that uncertainty reliably flags erroneous predictions.}
To further reduce overconfidence, we apply \textit{a-posteriori} temperature scaling to the trained models~\cite{guo2017calibrationmodernneuralnetworks}. 
For each model, we keep the weight fixed and use the validation logits and labels to tune a single scalar temperature~$T>0$: we perform a grid search over candidate values and select the optimal temperature $T^*$ that minimises the ECE on the validation set. At the test stage, the logits are divided by this learned temperature before applying the softmax,
to return calibrated class probabilities for each sample.

\section{Results}


Table~\ref{tab:conf_lidc_results_transposed} reports \add{calibration and classification performance on the LIDC-IDRI test set for all architectures and training modalities after a posteriori temperature scaling}. Baseline~\emph{(FL)} refers to standard focal-loss training without explicit uncertainty estimation, while \emph{MCD} and \emph{EDL} incorporate uncertainty estimation via stochastic dropout sampling and evidential learning, respectively. \add{In \emph{CA}, the focal loss augmented with a differentiable ECE surrogate, directly encourages well-calibrated predictions.} 
\add{Across all backbones, AUC values span roughly between ~$0.62$ and~$0.96$, with the most configurations clustering above~$0.90$, indicating that the majority of models retain strong discriminative ability on LIDC-IDRI.}
\add{Focal loss-based training frequently obtains the highest accuracy and F1 score for each architecture, whereas MCD and EDL variants remain close in classification performance and often return improved calibration (lower ECE and BS). }
\add{For several convolutional backbones (ResNet, DenseNet, and EfficientNet), the proposed uncertainty-aware training regimes achieve lower calibration error while maintaining strong classification performance. 
Moreover, in many cases, the proposed framework obtains calibration performance comparable to, or even better than, the calibration-aware objective, indicating that for certain architectures it provides clear improvements in both reliability and accuracy.}
\add{
MCD reduces ECE from $0.09$ to $0.04$ and BS from $0.22$ to $0.19$ while preserving AUC at $0.95$ and slightly improving accuracy to $0.89$. }
\add{In transformer-based backbones, the benefits of uncertainty-aware training are less consistent: for SwinViT3D and ViT3D, the baseline often remains competitive or superior in AUC and F1, and the uncertainty-aware variants mainly affect calibration rather than clearly improving overall performance.}

\begin{table}[t]
\centering
\caption{Calibration and classification performance on the LIDC-IDRI dataset after \textit{a posteriori} temperature scaling. Architectures are columns, and training regimes are rows. Bold: best per architecture. For ECE, MCE, and BS, lower is better ($\downarrow$); for AUC, F1, and Accuracy, higher is better ($\uparrow$).}
\label{tab:conf_lidc_results_transposed}
\small
\setlength{\tabcolsep}{3pt}
\resizebox{1\textwidth}{!}{%
\begin{tabular}{l|ccc|ccc|ccc|ccc|ccc|ccc|ccc}
\hline
& \multicolumn{3}{c|}{\textbf{DenseNet121}}
& \multicolumn{3}{c|}{\textbf{DenseNet169}}
& \multicolumn{3}{c|}{\textbf{ResNet18}}
& \multicolumn{3}{c|}{\textbf{ResNet50}}
& \multicolumn{3}{c|}{\textbf{EfficientNet3D}}
& \multicolumn{3}{c|}{\textbf{SwinViT3D}}
& \multicolumn{3}{c}{\textbf{ViT3D}} \\
\textbf{Method}
& ECE & MCE & BS
& ECE & MCE & BS
& ECE & MCE & BS
& ECE & MCE & BS
& ECE & MCE & BS
& ECE & MCE & BS
& ECE & MCE & BS \\
\hline
FL
& 0.06 & 0.32 & \textbf{0.15}
& 0.06 & 0.69 & 0.21
& 0.09 & 0.42 & 0.22
& 0.08 & 0.49 & 0.25
& 0.06 & 0.29 & \textbf{0.21}
& \textbf{0.04} & 0.16 & 0.32
& \textbf{0.04} & 0.10 & 0.38 \\
EDL
& 0.06 & 0.47 & 0.23
& \textbf{0.03} & \textbf{0.10} & \textbf{0.18}
& 0.08 & 0.75 & 0.24
& \textbf{0.07} & 0.67 & \textbf{0.24}
& 0.20 & 0.26 & 0.35
& 0.14 & 0.24 & 0.41
& 0.08 & 0.50 & \textbf{0.38} \\
MCD
& \textbf{0.05} & \textbf{0.16} & 0.20
& 0.09 & 0.27 & 0.22
& \textbf{0.04} & 0.47 & \textbf{0.19}
& 0.16 & \textbf{0.43} & 0.35
& \textbf{0.03} & \textbf{0.07} & 0.33
& 0.07 & \textbf{0.15} & \textbf{0.32}
& 0.10 & \textbf{0.10} & 0.50 \\
CA
& 0.11 & 0.53 & 0.21
& 0.08 & 0.15 & 0.17
& 0.07 & 0.41 & 0.19
& 0.14 & 0.85 & 0.27
& 0.19 & 0.88 & 0.37
& 0.08 & 0.39 & 0.34
& 0.08 & 0.15 & 0.38 \\
\hline
& \multicolumn{3}{c|}{\textbf{DenseNet121}}
& \multicolumn{3}{c|}{\textbf{DenseNet169}}
& \multicolumn{3}{c|}{\textbf{ResNet18}}
& \multicolumn{3}{c|}{\textbf{ResNet50}}
& \multicolumn{3}{c|}{\textbf{EfficientNet3D}}
& \multicolumn{3}{c|}{\textbf{SwinViT3D}}
& \multicolumn{3}{c}{\textbf{ViT3D}} \\
\textbf{Method}
& AUC & F1 & Acc
& AUC & F1 & Acc
& AUC & F1 & Acc
& AUC & F1 & Acc
& AUC & F1 & Acc
& AUC & F1 & Acc
& AUC & F1 & Acc \\
\hline
FL
& \textbf{0.95} & \textbf{0.91} & \textbf{0.91}
& \textbf{0.94} & 0.86 & 0.86
& 0.95 & 0.87 & 0.87
& \textbf{0.93} & 0.84 & 0.84
& \textbf{0.94} & \textbf{0.85} & \textbf{0.85}
& \textbf{0.85} & \textbf{0.76} & \textbf{0.76}
& \textbf{0.78} & 0.70 & 0.70 \\
EDL
& 0.92 & 0.86 & 0.86
& 0.90 & \textbf{0.90} & \textbf{0.90}
& 0.92 & 0.84 & 0.84
& 0.90 & \textbf{0.83} & 0.83
& 0.90 & 0.82 & 0.82
& 0.80 & 0.73 & 0.74
& 0.71 & \textbf{0.76} & \textbf{0.76} \\
MCD
& 0.94 & 0.89 & 0.89
& 0.93 & 0.83 & 0.83
& \textbf{0.95} & \textbf{0.88} & \textbf{0.89}
& 0.90 & 0.79 & \textbf{0.79}
& 0.92 & 0.74 & 0.75
& 0.84 & 0.73 & 0.74
& 0.62 & 0.51 & 0.60 \\
CA
& 0.89 & 0.88 & 0.89
& 0.93 & 0.91 & 0.91
& 0.96 & 0.89 & 0.89
& 0.93 & 0.84 & 0.84
& 0.91 & 0.80 & 0.80
& 0.84 & 0.75 & 0.75
& 0.78 & 0.71 & 0.71 \\
\hline
\end{tabular}%
}
\normalsize
\end{table}
\add{Table~\ref{tab:conf_ece_lidc} summarises how temperature scaling affects calibration across backbones and training regimes on LIDC-IDRI. For almost all configurations, a posteriori scaling substantially reduces ECE, with relative improvements typically in the range of $+25\%$ to $+60\%$.}
\add{Baseline models already benefit strongly from temperature scaling, achieving the largest relative ECE reductions for DenseNet169 ($+53\%$), ResNet50 ($+48\%$), EfficientNet3D ($+42\%$), SwinViT3D ($+46\%$), and ViT3D ($+28\%$). 
This confirms that a simple scalar temperature is highly effective even without explicit uncertainty-aware training.}
%
A small ECE increase in evidential training suggests that some architectures are already well calibrated before temperature scaling. 
\add{However, in line with Ref.~\cite{chen2024revisitingessentialnonessentialsettings}, this behaviour may also indicate a tendency towards overconfident logits.}

\begin{figure}[t]
    \centering
    \begin{tabular}{cc}
         \includegraphics[width=0.4\textwidth]{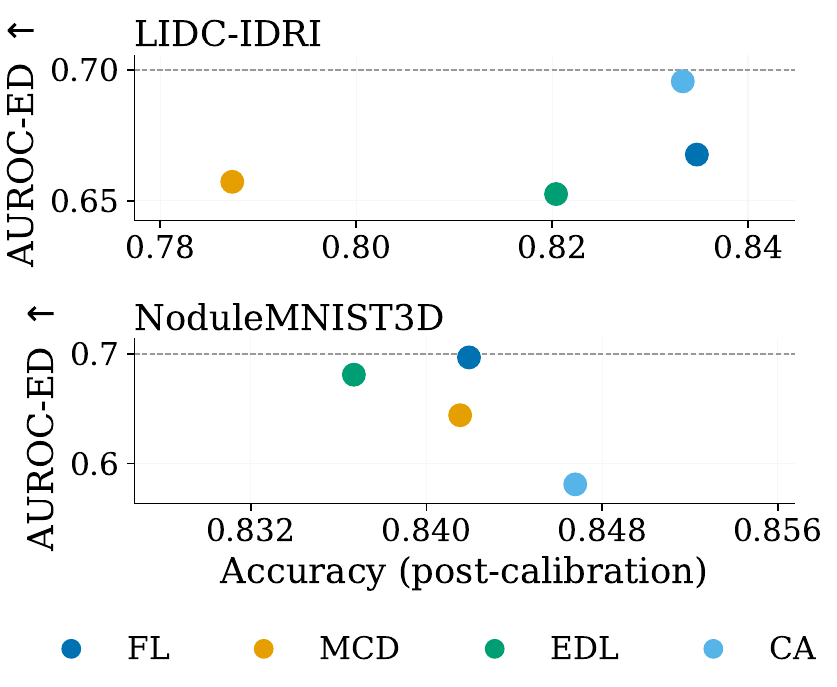}&  \hspace{0.2cm}
         \includegraphics[width=0.55\textwidth]{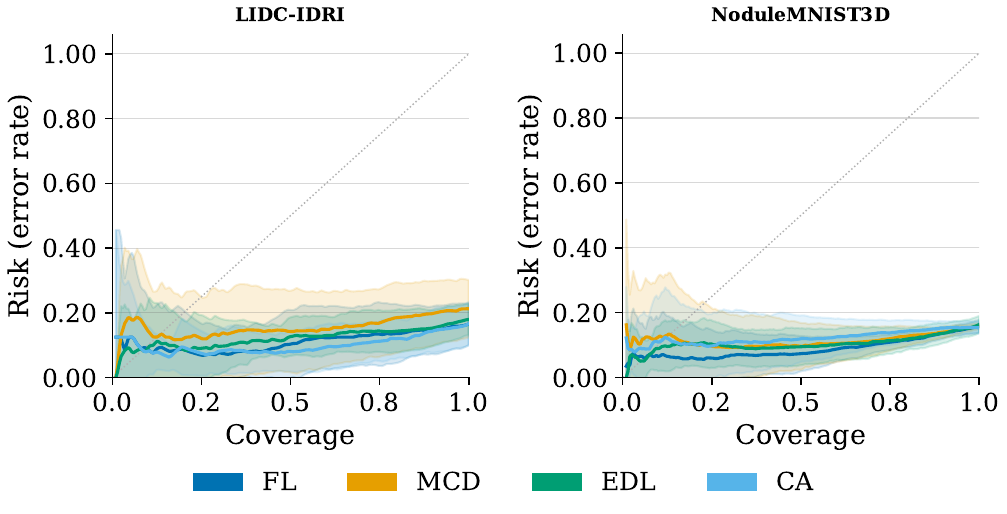} \\
         {\small (a)} & {\small (b)} \\
    \end{tabular}

    \caption{%
    (a) Post-calibration classification accuracy vs.\ uncertainty quality (AUROC-ED) for each training mode across the two datasets. Each point represents the mean over all evaluated architectures. The dashed line marks AUROC-ED = 0.70, a reference threshold for practically useful error detection; points in the upper-right quadrant achieve both high accuracy and reliable uncertainty estimates.
    (b) Risk–coverage curves for selective prediction across the two datasets. Risk (error rate) is measured on the fraction of predictions retained after abstaining on the most uncertain samples (coverage). Solid lines show the mean across architectures; shaded bands show $\pm 1$ standard deviation. Lower curves indicate better uncertainty-guided abstention, where the model concentrates errors in the samples it chooses to defer.}
    \label{fig:uq_risk_coverage}
\end{figure}

\begin{table}[t]
\centering
\caption{ECE before and after temperature scaling on LIDC-IDRI. $\Delta$ denotes relative improvement (\%) after scaling.}
\label{tab:conf_ece_lidc}
\footnotesize
\setlength{\tabcolsep}{4pt}
\resizebox{1\textwidth}{!}{%
\begin{tabular}{l|ccc|ccc|ccc|ccc|ccc|ccc|ccc}
\hline
& \multicolumn{3}{c|}{\textbf{DenseNet121}} 
& \multicolumn{3}{c|}{\textbf{DenseNet169}}
& \multicolumn{3}{c|}{\textbf{ResNet18}}
& \multicolumn{3}{c|}{\textbf{ResNet50}}
& \multicolumn{3}{c|}{\textbf{EfficientNet3D}}
& \multicolumn{3}{c|}{\textbf{SwinViT3D}}
& \multicolumn{3}{c}{\textbf{ViT3D}} \\
\textbf{Method} 
& ECE\(_\text{b}\) & ECE\(_\text{a}\) & $\Delta$ 
& ECE\(_\text{b}\) & ECE\(_\text{a}\) & $\Delta$
& ECE\(_\text{b}\) & ECE\(_\text{a}\) & $\Delta$
& ECE\(_\text{b}\) & ECE\(_\text{a}\) & $\Delta$
& ECE\(_\text{b}\) & ECE\(_\text{a}\) & $\Delta$
& ECE\(_\text{b}\) & ECE\(_\text{a}\) & $\Delta$
& ECE\(_\text{b}\) & ECE\(_\text{a}\) & $\Delta$ \\
\hline
Baseline 
& 0.087 & 0.061 & +29
& 0.118 & 0.056 & \textbf{+53}
& 0.128 & 0.087 & +32
& 0.160 & 0.083 & \textbf{+48}
& 0.099 & 0.058 & \textbf{+42}
& 0.072 & \textbf{0.039} & \textbf{+46}
& 0.061 & \textbf{0.044} & \textbf{+28} \\
EDL
& 0.089 & 0.060 & +32
& 0.046 & \textbf{0.027} & +40
& 0.130 & 0.075 & +42
& 0.065 & \textbf{0.065} & 0
& 0.256 & 0.201 & +22
& 0.112 & 0.143 & -28
& 0.096 & 0.080 & +16 \\
MCD
& 0.119 & \textbf{0.052} & \textbf{+56}
& 0.107 & 0.093 & +14
& 0.106 & \textbf{0.043} & \textbf{+59}
& 0.203 & 0.163 & +20
& 0.039 & \textbf{0.029} & +25
& 0.128 & 0.072 & +44
& 0.098 & 0.098 & 0 \\
CA
& 0.109 & 0.109 & -1
& 0.091 & 0.081 & +11
& 0.119 & 0.074 & +37
& 0.155 & 0.140 & +10
& 0.184 & 0.186 & -1
& 0.070 & 0.085 & -22
& 0.103 & 0.076 & +26 \\
\hline
\end{tabular}%
}
\normalsize
\end{table}
To assess generalisation, we evaluate all methods on NoduleMNIST3D~\cite{Yang_2023} in Table~\ref{tab:conf_nodulemnist_results_transposed}. 
Across all architectures, AUC and F1 are comparable to LIDC-IDRI (Table~\ref{tab:conf_lidc_results_transposed}), indicating that our training strategies transfer well to this dataset.
\add{Classification performance is mixed across backbones: the focal-loss baseline leads for DenseNet121, EfficientNet3D, and ViT3D, whereas EDL achieves the best F1 for DenseNet169 and ResNet18, with MCD performing competitively throughout.}
\add{The best calibration (lowest ECE) is frequently obtained by uncertainty-driven variants, particularly DenseNet169 with evidential training.}
In contrast, DenseNet121 with EDL shows some degraded performance (AUC $= 0.81$ vs.\ $0.83$ for MCD). \add{The clearest case of EDL degradation occurs on ViT3D, where evidential training collapses to AUC $= 0.50$ and F1 $= 0.70$ compared to $0.83$ and $0.83$, respectively, for the baseline.}
\add{This suggests that the Dirichlet evidence parameterisation may be incompatible with the attention-based architecture on this dataset and that lower-resolution volumes may require more careful hyperparameter tuning.}
Overall, uncertainty-driven training maintains competitive classification performance and improves or matches calibration on NoduleMNIST3D, in line with the observations \add{reported} by Zahari~\emph{et al.}~\cite{Zahari2023}.
%

\add{Figure~\ref{fig:uq_risk_coverage}(a) plots post-calibration accuracy against AUROC-ED for each training mode, revealing a dataset-dependent trade-off between classification performance and uncertainty quality. On LIDC-IDRI, CA training occupies the top-right region with the highest AUROC-ED ($0.696$) and near-best accuracy ($0.833$), just crossing the $0.70$ practical utility threshold, while FL achieves a similar accuracy ($0.835$) with slightly weaker uncertainty separation~($0.668$). On NoduleMNIST, the picture reverses: FL obtains the best uncertainty quality ($0.697$, just below the threshold), whereas CA training yields the highest accuracy ($0.847$) but the lowest AUROC-ED ($0.581$) of all modes, indicating that directly minimising calibration error does not automatically produce a useful error-detection signal. 
MCD consistently underperforms in uncertainty quality across both datasets, despite competitive accuracy.
}
\add{Figure~\ref{fig:uq_risk_coverage}(b) contextualises these scalar metrics through full risk–coverage curves. On LIDC-IDRI, FL and CA training maintain the lowest and most stable risk across coverage levels, while MCD shows the widest uncertainty band, reflecting high inter-architecture variability in its uncertainty estimates. On NoduleMNIST3D, all four training methods collapse into a tight band with similarly low risk at full coverage.}
Training-time overhead for evidential models compared to the deterministic baselines was found to be modest. In particular, for DenseNet121 and DenseNet169, evidential training required only about~$20-30\%$ more time per fold, and for the ResNet architectures, the overhead was below~$15\%$. 
In contrast, MCD training was substantially more expensive, increasing training time by roughly $100-150\%$ relative to the baseline for the DenseNet models and by about $50-70\%$ for the ResNets.
In terms of the inference time, evidential models were more efficient than MCD; evidential inference was typically $70-80\%$ faster than MCD for all architectures, and about $75-85\%$ faster than the baseline for the DenseNets and $40-60\%$ faster for the ResNets.
Thus, evidential models improve calibration while adding only a small training overhead and achieving large relative savings in inference time compared to MCD.

\begin{table}[t]
\centering
\caption{Calibration and classification performance on the NoduleMNIST3D dataset after \textit{a posteiori} temperature scaling. Architectures are columns and training regimes are rows. Bold: best per architecture. For ECE, MCE, and BS, lower is better ($\downarrow$); for AUC, F1, and Accuracy, higher is better ($\uparrow$).}
\label{tab:conf_nodulemnist_results_transposed}
\small
\setlength{\tabcolsep}{3pt}
\resizebox{1\textwidth}{!}{%
\begin{tabular}{l|ccc|ccc|ccc|ccc|ccc|ccc|ccc}
\hline
& \multicolumn{3}{c|}{\textbf{DenseNet121}}
& \multicolumn{3}{c|}{\textbf{DenseNet169}}
& \multicolumn{3}{c|}{\textbf{ResNet18}}
& \multicolumn{3}{c|}{\textbf{ResNet50}}
& \multicolumn{3}{c|}{\textbf{EfficientNet3D}}
& \multicolumn{3}{c|}{\textbf{SwinViT3D}}
& \multicolumn{3}{c}{\textbf{ViT3D}} \\
\textbf{Method}
& ECE & MCE & BS
& ECE & MCE & BS
& ECE & MCE & BS
& ECE & MCE & BS
& ECE & MCE & BS
& ECE & MCE & BS
& ECE & MCE & BS \\
\hline
Baseline
& \textbf{0.07} & 0.16 & \textbf{0.23}
& 0.06 & \textbf{0.14} & 0.25
& 0.05 & \textbf{0.08} & 0.25
& \textbf{0.05} & \textbf{0.11} & \textbf{0.23}
& 0.05 & 0.30 & \textbf{0.21}
& \textbf{0.04} & \textbf{0.12} & 0.22
& 0.05 & 0.15 & \textbf{0.25} \\
EDL
& 0.10 & 0.31 & 0.27
& \textbf{0.04} & 0.24 & 0.25
& 0.06 & 0.15 & \textbf{0.21}
& 0.10 & 0.54 & 0.24
& 0.07 & 0.27 & 0.25
& 0.06 & 0.17 & \textbf{0.22}
& 0.08 & \textbf{0.08} & 0.34 \\
MCD
& 0.08 & \textbf{0.15} & 0.24
& 0.06 & 0.17 & \textbf{0.24}
& \textbf{0.03} & 0.16 & 0.22
& 0.09 & 0.22 & 0.26
& \textbf{0.03} & \textbf{0.08} & 0.22
& 0.05 & 0.18 & 0.24
& \textbf{0.04} & 0.08 & 0.25 \\
\hline
& \multicolumn{3}{c|}{\textbf{DenseNet121}}
& \multicolumn{3}{c|}{\textbf{DenseNet169}}
& \multicolumn{3}{c|}{\textbf{ResNet18}}
& \multicolumn{3}{c|}{\textbf{ResNet50}}
& \multicolumn{3}{c|}{\textbf{EfficientNet3D}}
& \multicolumn{3}{c|}{\textbf{SwinViT3D}}
& \multicolumn{3}{c}{\textbf{ViT3D}} \\
\textbf{Method}
& AUC & F1 & Acc
& AUC & F1 & Acc
& AUC & F1 & Acc
& AUC & F1 & Acc
& AUC & F1 & Acc
& AUC & F1 & Acc
& AUC & F1 & Acc \\
\hline
Baseline
& \textbf{0.85} & \textbf{0.86} & \textbf{0.86}
& 0.83 & 0.84 & 0.84
& 0.86 & 0.82 & 0.82
& 0.81 & \textbf{0.83} & 0.84
& \textbf{0.89} & \textbf{0.87} & \textbf{0.87}
& 0.85 & \textbf{0.84} & \textbf{0.85}
& \textbf{0.83} & \textbf{0.83} & \textbf{0.83} \\
EDL
& 0.81 & 0.83 & 0.84
& \textbf{0.84} & \textbf{0.86} & \textbf{0.85}
& \textbf{0.88} & \textbf{0.87} & \textbf{0.87}
& \textbf{0.87} & 0.83 & \textbf{0.85}
& 0.83 & 0.85 & 0.85
& 0.82 & 0.84 & 0.85
& 0.50 & 0.70 & 0.79 \\
MCD
& 0.83 & 0.85 & 0.85
& 0.83 & 0.83 & 0.84
& 0.85 & 0.85 & 0.85
& 0.79 & 0.83 & 0.84
& 0.85 & 0.85 & 0.85
& \textbf{0.85} & 0.84 & 0.84
& 0.80 & 0.81 & 0.83 \\
\hline
\end{tabular}%
}
\normalsize
\end{table}

\section{Conclusions}
We introduced an uncertainty-driven training framework for 3D lung nodule classification that integrates uncertainty quantification directly into the training stage. 
By using validation-based uncertainty to dynamically reweight classes during training, the framework explicitly targets underperforming and highly uncertain classes, addressing both class imbalance and miscalibration.
Across seven backbone architectures \add{and three uncertainty-aware training modes (Monte Carlo Dropout (MCD), Evidential Deep Learning (EDL), and calibration-aware training (CA)}, experiments on LIDC-IDRI dataset show that uncertainty-driven training maintains competitive discrimination (AUC up to $0.95$ and F1 up to $0.91$ while improving calibration. 
\add{MCD yields the strongest calibration gains on convolutional backbones, achieving ECE reductions of roughly $56\%$ for DenseNet121 and $59\%$ for ResNet18 while preserving AUC above $0.94$ and $0.95$, respectively. EDL achieves the best post-scaling calibration on DenseNet169 (ECE $= 0.027$), but is less consistent on ResNet50 and attention-based architectures.} 
\add{Across architectural families, uncertainty-driven training yields more consistent gains on convolutional backbones than on transformer-based architectures, where the interaction between attention and the uncertainty objective appears more sensitive to dataset resolution and scale.}
Calibration metrics and temperature scaling analyses indicate that uncertainty-aware models are less overconfident and can be further refined with calibration when needed.
\add{This supports that a simple scalar temperature can be highly effective even in the absence of explicit uncertainty-aware training.}
Validation on the NoduleMNIST3D benchmark demonstrates that these conclusions generalise beyond a single clinical dataset, as uncertainty-driven variants retain strong AUC and~F1 scores while often matching or improving calibration compared to their baselines. 
\add{A notable exception is ViT3D with EDL, for which AUC and F1 drop to~$0.50$ and~$0.70$, respectively, on NoduleMNIST3D. This suggests that the Dirichlet evidence parameterisation can be incompatible with attention-based architectures under the lower-resolution volumes of this dataset.}

Our results suggest that incorporating class-level uncertainty into the training loop represents a significant practical step towards more reliable and clinically trustworthy 3D medical imaging models. 
These gains come at a modest computational cost: the evidential training is close to the baseline in training time and substantially faster at inference, while MCD roughly doubles the total runtime due to multiple stochastic passes, it remains feasible in practice.
Future works will extend this framework to multi-class nodule subtype classification and 3D segmentation, where pixel- or voxel-level uncertainty could guide region-of-interest prioritisation. 
Given the strong performance of evidential learning on several convolutional backbones, it will also be important to explore architectural and training variants that improve EDL stability on transformer-based models, and to assess whether the resulting uncertainty estimates remain well calibrated at scale.
\add{This study has several limitations that motivate future investigation. First, evaluation is restricted to binary malignancy classification (LIDC-IDRI) and a two-class nodule detection task (NoduleMNIST3D); further validation on multi-class or multi-label problems is needed to confirm the generalisability of the calibration gains. 
Second, our uncertainty reweighting operates at the class level over the entire dataset and does not adapt weights based on individual samples or spatial regions within an image. This coarse-grained approach may be less effective for tasks with fine-grained inter-class structure, where local or sample-specific uncertainty would be more informative.
Third, the degraded performance of EDL on transformer-based architectures (for example, ViT3D on NoduleMNIST3D) remains poorly understood and requires additional examination.
Finally, MCD requires $K = 20$ stochastic forward passes at inference, which may constrain real-time deployment. However, distillation or amortised-inference approaches could mitigate this overhead without sacrificing uncertainty quality.}

\begin{credits}
\subsubsection{\ackname} 
The authors gratefully acknowledge Dr. Mingfei Sun from The University of Manchester (UoM) for his insightful comments and guidance on this work. They also acknowledge the use of the Computational Shared Facility (CSF) at UoM.
\end{credits}

%
%
%
\bibliographystyle{splncs04}
\bibliography{mybibliography}
%
%
%
%
%
\end{document}